\documentclass[%
 aip,
 apl,
superscriptaddress
amsmath,amssymb,
reprint,%
]{revtex4-1}

\usepackage{graphicx}% Include figure files
\usepackage{dcolumn}% Align table columns on decimal point
\usepackage{bm}% bold math
\usepackage{amsmath} % for align environment
\usepackage[utf8]{inputenc}
\usepackage[T1]{fontenc}
\usepackage{mathptmx}
\usepackage[english]{babel}
\usepackage{etoolbox}

\usepackage{xcolor}
\usepackage{etoolbox}

\usepackage[hidelinks]{hyperref}
\usepackage{orcidlink}  

\usepackage{xcolor}
\usepackage{etoolbox}

\newif\ifcolorcite
\colorcitetrue  % colorcitefalse->black

\let\oldcite\cite
\renewcommand{\cite}[1]{%
  \ifcolorcite
    {\color{blue}\oldcite{#1}}%
  \else
    \oldcite{#1}%
  \fi
}

\let\oldref\ref
\renewcommand{\ref}[1]{%
  \ifcolorcite
    {\color{blue}\oldref{#1}}%
  \else
    \oldref{#1}%
  \fi
}
\makeatletter
\def\@email#1#2{%
 \endgroup
 \patchcmd{\titleblock@produce}
  {\frontmatter@RRAPformat}
  {\frontmatter@RRAPformat{\produce@RRAP{*#1\href{mailto:#2}{#2}}}\frontmatter@RRAPformat}
  {}{}
}%
\makeatother
\begin{document}

%\preprint{Prepared for APL}

\title[]{Optimization of EUV output by experimentally validated radiation-hydrodynamic simulations across a broad laser parameter space}
% Force line breaks with \\

\author{Nozomi Tanaka \orcidlink{0000-0002-8392-2224}}
\email{tanaka.nozomi.ile@osaka-u.ac.jp}
% \thanks{N. Tanaka and Y. Yamamoto contributed equally to this work.}
\affiliation{ 
Institute of Laser Engineering, The University of Osaka, 2-6 Yamadaoka, Suita, Osaka 565-0871, Japan%\\This line break forced with \textbackslash\textbackslash
}%
\affiliation{
National Institute for Fusion Science, 322-6 Oroshi, Toki, Gifu 509-5292, Japan%\\This line break forced with \textbackslash\textbackslash
}

\author{Yu Yamamoto \orcidlink{0000-0002-1549-6382}}
% \thanks{N. Tanaka and Y. Yamamoto contributed equally to this work.}
\affiliation{ 
Institute of Laser Engineering, The University of Osaka, 2-6 Yamadaoka, Suita, Osaka 565-0871, Japan%\\This line break forced with \textbackslash\textbackslash
}%

\author{Akira Sasaki \orcidlink{0000-0002-5094-7361}}
\affiliation{
Kansai Institute for Photon Science, National Institutes for Quantum Science and Technology, 8-1-7 Umemidai, Kizugawa, Kyoto 619-0215, Japan%\\This line break forced with \textbackslash\textbackslash
}%
\affiliation{ 
Institute of Laser Engineering, The University of Osaka, 2-6 Yamadaoka, Suita, Osaka 565-0871, Japan%\\This line break forced with \textbackslash\textbackslash
}%

\author{Katsunobu Nishihara \orcidlink{0000-0002-5572-1169}}
\affiliation{ 
Institute of Laser Engineering, The University of Osaka, 2-6 Yamadaoka, Suita, Osaka 565-0871, Japan%\\This line break forced with \textbackslash\textbackslash
}%
\affiliation{Faculty of Engineering, Osaka Metropolitan University, 3-3-138 Sugimoto, Sumiyoshi-ku, Osaka 558-8585, Japan}

\author{Atsushi Sunahara \orcidlink{0000-0001-7543-5226}}
\affiliation{%
Center for Materials Under eXtreme Environment, (CMUXE), School of Nuclear Engineering, Purdue University, 500 Central Drive, West Lafayette, Indiana 47907, USA%\\This line break forced% with \\
}%
\affiliation{ 
Institute of Laser Engineering, The University of Osaka, 2-6 Yamadaoka, Suita, Osaka 565-0871, Japan%\\This line break forced with \textbackslash\textbackslash
}%

\author{Tomoyuki Johzaki \orcidlink{0000-0002-5738-4661}}
\affiliation{%
Graduate School of Advanced Science and Engineering, Hiroshima University, 1-4-1 Kagamiyama, Higashihiroshima, Hiroshima 739-8527, Japan%\\This line break forced% with \\
}%
\affiliation{ 
Institute of Laser Engineering, The University of Osaka, 2-6 Yamadaoka, Suita, Osaka 565-0871, Japan%\\This line break forced with \textbackslash\textbackslash
}%

\author{Yuji Takagi \orcidlink{0000-0002-0922-2318}}
\affiliation{ 
Institute of Laser Engineering, The University of Osaka, 2-6 Yamadaoka, Suita, Osaka 565-0871, Japan%\\This line break forced with \textbackslash\textbackslash
}%

\author{Kentaro Tomita \orcidlink{0000-0002-0748-7953}}
\affiliation{ 
Graduate School of Engineering, Hokkaido University, Kita 13, Nishi 8, Kita-ku, Sapporo, Hokkaido 060-8628, Japan%\\This line break forced with \textbackslash\textbackslash
}%

\author{Shinsuke Fujioka \orcidlink{0000-0001-8406-1772}}
\affiliation{ 
Institute of Laser Engineering, The University of Osaka, 2-6 Yamadaoka, Suita, Osaka 565-0871, Japan%\\This line break forced with \textbackslash\textbackslash
}%

\author{Masashi Yoshimura \orcidlink{0000-0002-0770-1251}}
\affiliation{ 
Institute of Laser Engineering, The University of Osaka, 2-6 Yamadaoka, Suita, Osaka 565-0871, Japan%\\This line break forced with \textbackslash\textbackslash
}%

\date{\today}% It is always \today, today,
             %  but any date may be explicitly specified

\begin{abstract}
Practical requirements such as improving wall-plug efficiency and reducing system footprint have become increasingly important with the introduction of extreme ultraviolet (EUV) lithography into high-volume semiconductor manufacturing.
These demands motivate the development of solid-state mid-infrared lasers as alternatives to current CO$_2$ lasers.
Systematic exploration of laser-to-EUV conversion efficiency (EUV-CE) over a broad parameter space is essential when altering the drive laser's wavelength, because the EUV-CE depends on the laser parameters in a complex manner.
In this work, we performed a large-scale grid search of more than 140,000 parameter combinations for laser-produced tin plasma EUV sources using the radiation-hydrodynamics code \textsc{STAR-1D}, which is validated against EUV source experiments.
The systematic wavelength dependence of the optimum pulse width and target size is governed by the requirement to simultaneously achieve the electron temperature and density optimal for EUV emission, maintain efficient laser absorption, and suppress EUV self-absorption.
The resulting CE map predicts a global maximum of 5.63\% at 5.5~$\mu$m.
For the practically relevant 2~$\mu$m solid-state driver, a maximum CE of 4.64\% is obtained, in good agreement with recent experimental results. Multiple operating points are identified over a broad range of pulse parameters, providing guidance for 2~$\mu$m-driven EUV source development.
\end{abstract}

\maketitle

%\clearpage

Extreme ultraviolet (EUV) lithography has become the backbone of advanced semiconductor manufacturing, relying on laser-produced plasma sources in which a high-power CO$_2$ laser irradiates pre-shaped tin micro-droplet targets to generate 13.5~nm EUV light.\cite{Versolato2019,Versolato2022}
As the technology has matured, the challenges of scaling EUV source power have increasingly become not only a plasma physics problem but also a systems engineering one, where the electrical power efficiency, compactness, and reliability of the entire laser driver system are critical considerations.
While CO$_2$ laser-driven sources currently achieve laser-to-EUV conversion efficiencies (EUV-CE or CE) of approximately 5--6\%,\cite{Mostafa2023} recent advances in mid-infrared solid-state lasers have been proposed as a viable alternative drivers owing to their higher wall-plug efficiencies and more compact form factors.\cite{Siders2019,Versolato2022}
Here, the EUV-CE is defined as the ratio of the EUV energy emitted into a 2$\pi \ \mathrm{sr}$ solid angle within a 2\% wavelength-bandwidth centered at 13.5~nm to the incident laser energy.

The optimal operating parameters for such solid-state driver lasers are not immediately obvious: Energy of the shorter wavelength is absorbed in a higher electron density region
%($n_\mathrm{c} \propto \lambda^{-2}$) 
compared to CO$_2$ lasers, necessitating a re-optimization of the driver parameters to reach efficient EUV emission.
Among the candidate wavelengths, 2~$\mu$m lasers have attracted the most attention as potential alternatives to CO$_2$ lasers for EUV driver, benefiting from technologies originally developed for high-harmonic generation and molecular spectroscopy.\cite{Bock2025,Heuermann2022,Tomilov2022}
However, existing simulations\cite{Hemminga2023,Dong2025} and experiments\cite{Mostafa2023} have not yet provided a comprehensive mapping of EUV-CEs in the parameter space across the mid-infrared wavelength range.
The use of a well-validated one-dimensional radiation-hydrodynamics code enables such a survey with confidence, allowing a dense and reliable scan over a broad parameter space at manageable computational cost.
We therefore performed such a survey using STAR-1D,\cite{Sunahara2008,Nishihara2008} revealing the wavelength-dependent scaling of optimum laser and target conditions and identifying broad practical operating windows for mid-infrared-driven EUV sources.

\textsc{STAR-1D} code has been extensively validated against laboratory experiments; in particular, the computed EUV-CE and spectral characteristics were shown to be in good agreement with measurements from uniform spherical-target irradiation experiments on the GEKKO XII laser.\cite{Shimada2005}
Furthermore, the opacity tables used in \textsc{STAR-1D} code, computed using a detailed collisional-radiative model,\cite{Sasaki2010} have recently been extended to cover charge states up to Sn$^{17+}$, improving the reliability of predictions at higher laser intensities where higher ionization states become significant.\cite{Sasaki2026}

A large-scale grid search was performed, simulating more than 140,000 of different combinations of input parameters.
The simulations assume a spherical tin target with a uniform initial mass density of $\rho_0 = 0.04$~g/cm$^3$,\cite{Langer2019} fixed across all parameter combinations, intended to represent a rarefied tin target prior to main-pulse irradiation.
Note that the one-dimensional spherical approximation is valid when the plasma scale length is small enough compared to the initial target diameter.
The variable input parameters were: laser intensity ($1 \times 10^{9}$--$1 \times 10^{11}$~W/cm$^{2}$), wavelength (1--11~$\mu$m), pulse width (0.5, 1.0, 10.0, and 30.0~ns for all models, with an additional 100~ns case for the 2~$\mu$m wavelength), target radius (5, 50, 100, 200, 400, 600, 800, and 1000~$\mu$m), and pulse shape, characterized by the half width at the half maximum (HWHM) ratio of the rising and falling edge (1:9 or 1:1) for super-Gaussian pulses of $2$nd or $10$th order applied before and after the pulse peak.
The laser spatial profile was modeled as a 10th-order super-Gaussian beam centered on the target with an focusing optics $F$-number of 3, where the focal spot diameter was set a half of the target diameter.

\begin{figure*}
\centering
    \includegraphics[width=0.8\textwidth]{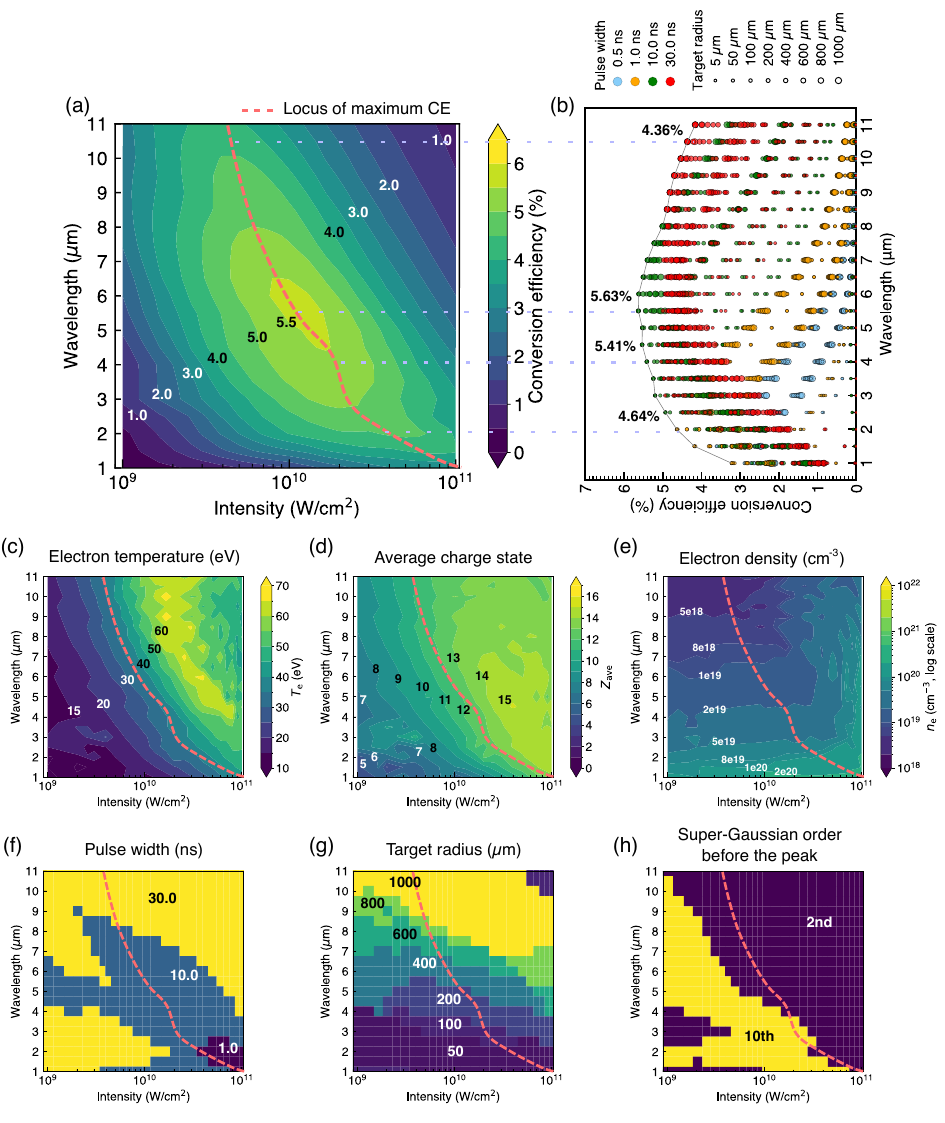}
\caption{(a) Contour map of the EUV conversion efficiency (EUV-CE).
The red dashed line indicates the locus of the maximum CE at each wavelength.
(b) Distribution of maximum CE values at each wavelength along the locus in panel (a). 
The black envelope curve indicates the maximum CE at each wavelength.
Panels (c)--(h) show plasma and laser parameters for the maximum CE models in panel (a): (c) electron temperature, (d) average charge state, and (e) electron density at the opacity-weighted emission position (emissivity-weighted mean position accounting for optical depth); (f) laser pulse width (ns), (g) target radius ($\mu$m), and (h) super-Gaussian order before the pulse peak.}
\label{fig:ce_map}
\end{figure*}

\begin{figure*}
\centering
    \includegraphics[width=1.0\textwidth]{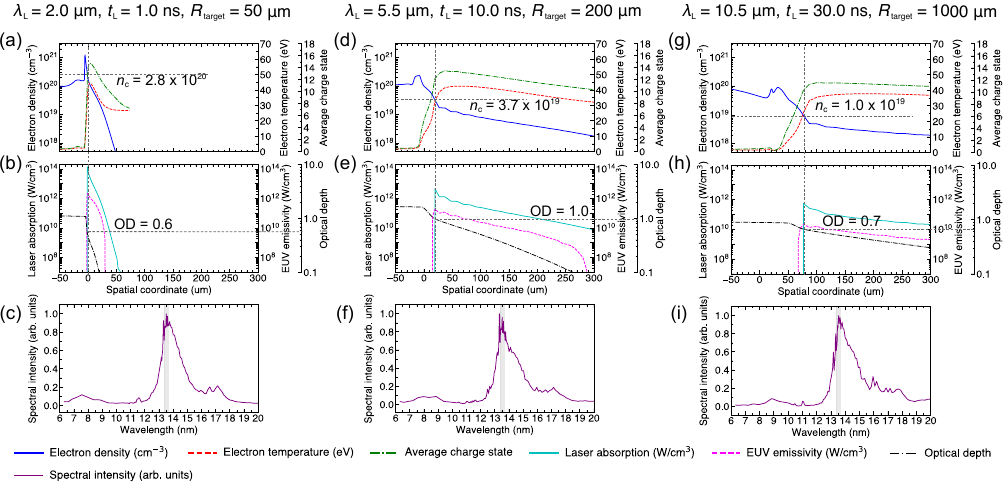}
\caption{Plasma and emission profiles at the time of the laser pulse peak for the conditions corresponding to each wavelength of along the locus in Fig.~\ref{fig:ce_map}(a).
Spatial profiles of (a,d,g) electron density (blue solid line), electron temperature (red dashed line), and average charge state (green dash-dotted line); and (b,e,h) laser absorption (turquoise solid line), EUV in-band emissivity (magenta dashed line), and optical depth (black dash-dotted line).
The spatial coordinate 0 corresponds to the initial position of the target surface.
(c,f,i) Time-integrated EUV spectra under the same conditions, where the hatched area indicates the in-band region within the 2\% bandwidth centered at $13.5~\mathrm{nm}$.
The laser wavelengths are (a--c) $2.0~\mu\mathrm{m}$, (d--f) $5.5~\mu\mathrm{m}$, and (g--i) $10.5~\mu\mathrm{m}$, at laser intensities of $5.0 \times 10^{10}~\mathrm{W/cm^2}$, $1.29 \times 10^{10}~\mathrm{W/cm^2}$, and $4.0 \times 10^{9}~\mathrm{W/cm^2}$, respectively.
The peak position of EUV emissivity, critical densities $n_{\mathrm{c}}$ for each wavelength, and the optical depth at the position of peak EUV emissivity are also indicated with gray dashed lines.
}
\label{fig:profile}
\end{figure*}

Figure~\ref{fig:ce_map}(a) presents a contour map of CE in the driver laser intensity--wavelength space, where each grid point represents the maximum over 256 simulations.
The red dashed line denotes the locus of the maximum CE for each wavelength.
Figure~\ref{fig:ce_map}(b) shows the distribution of CE values at each wavelength along the locus shown in panel (a), and the black envelope curve indicates the maximum CE at each wavelength.
The highest CE obtained in this survey is 5.63\% at a wavelength of 5.5~$\mu$m, while wavelengths of 2~$\mu$m and 4~$\mu$m yield CE of 4.64\% and 5.41\%, respectively. These values exceed the 4.36\% obtained at 10.5~$\mu$m, which is nearly the wavelength of the current CO$_2$ main pulse laser, highlighting the potential of shorter-wavelength drivers for EUV generation.

The wavelength dependence of the optimum laser intensity has been discussed in previous studies of Sn laser-produced plasmas.
Nishihara \textit{et al.} proposed, based on a power-balance model, that the optimum laser intensity $I_{\mathrm{L}}^{\mathrm{opt}}$ decreases with increasing laser wavelength $\lambda$ ($\mathrm{\mu m}$), approximately following $I_{\mathrm{L}}^{\mathrm{opt}} \propto \lambda^{-1.2}$.\cite{Nishihara2008}
This trend is also consistent with experimental observations for 1--2~$\mu$m laser-driven Sn plasmas, where similar EUV spectra and charge-state distributions were obtained when the drive-laser intensity was reduced at the longer wavelength.\cite{Schupp2019,Behnke2021,Schupp2021a,Versolato2022}
Hemminga \textit{et al.} performed simulations using an experimentally motivated intensity scaling of $I_{\mathrm{L}}^{\mathrm{opt}} \propto \lambda^{-1}$,\cite{Hemminga2023,Nishihara2006} while the simulations by Dong \textit{et al.} yielded a scaling of $I_{\mathrm{L}}^{\mathrm{opt}} \propto \lambda^{-1.2}$,\cite{Dong2025} both being broadly consistent with the Nishihara's scaling.
In the present parameter survey, the CE map in the wavelength--intensity space exhibits a similar systematic trend.
By extracting the locus of the maximum CE at each wavelength in Fig.~\ref{fig:ce_map}(a), we obtain

\begin{equation}
    I_{\mathrm{L}}^{\mathrm{opt}} = 1.1 \times 10^{11} \cdot (\lambda)^{-1.36}.
\label{eq:scale}
\end{equation}

\noindent This scaling is consistent with the previously established physical picture.
For comparison, the CE locus reported by LLNL group shows $I_{\mathrm{L}}^{\mathrm{opt}} \propto \lambda^{-0.4}$.\cite{Siders2019}
This scaling is substantially flatter than those reported in previous studies and is inconsistent with both the results of the present work and those of earlier studies.

Figure~\ref{fig:ce_map}(c), (d), and (e) show the electron temperature, 
average charge state, and electron density for the maximum CE conditions 
in panel~(a), evaluated at the laser pulse peak and the escaping-emission-weighted position $\langle r \rangle_{\mathrm{em}}$ weighted by $\epsilon_\nu(r)\,e^{-\tau_\nu(r)}$, where $\epsilon_\nu(r)$
is the local emissivity and $e^{-\tau_\nu(r)}$ is the attenuation factor
representing the local emission that escapes the plasma and reaches the observer.
Here the optical depth is defined as

\begin{equation}
\tau_\nu(r) = \int_{r}^{\infty} \kappa_\nu(r')\, dr',
\label{eq:emission_position}
\end{equation}

\noindent where $\kappa_\nu(r)$ is the local opacity.
In panels (c) and (d), the locus runs through regions of electron temperature $\sim$35~eV and average charge states of 11--13, confirming that the maximum CE conditions correspond to the optimum plasma conditions for in-band EUV emission\cite{Sasaki2007,Nishihara2008}.
Panel (e) shows that the electron density at this position remains close to the critical density across all wavelengths $n_{\mathrm{c}}\,[\mathrm{cm^{-3}}] = 1.1 \times 10^{21} / (\lambda\,[\mu\mathrm{m}])^{2}$.
These results demonstrate that, across the full range of driver wavelengths investigated, optimal plasma conditions for efficient laser absorption and in-band EUV emission can be simultaneously achieved.

Figure~\ref{fig:ce_map}(f) and (g) represent the optimum laser pulse width (ns) and target radius ($\mu$m).
As seen in Fig.~\ref{fig:ce_map}(f) and (g), shorter driver wavelengths require both shorter pulse widths and smaller target radii to achieve high CE indicated by the locus line.
For example, at the long-wavelength end (10.5~$\mu$m), the optimum parameters are a pulse width of 30.0~ns, a target radius of 1000~$\mu$m, and an intensity of $4.0 \times 10^{9}$~W/cm$^{2}$, whereas at the short-wavelength end (2.0~$\mu$m), they shift to 1.0~ns, 50~$\mu$m, and $5.0 \times 10^{10}$~W/cm$^{2}$, respectively.

These trend are explained as follows.
First, the optimal laser intensity is set by the wavelength-dependent laser absorption condition.
At longer wavelengths, the lower critical density places the laser absorption in a lower-density plasma, where a moderate laser intensity suffices to heat the plasma to the optimal EUV-emitting temperature.
Excessively high intensities drive the outer low-density region into a high-temperature regime as shown in Fig.~\ref{fig:ce_map}(c), where the laser absorption coefficient $\kappa_a \propto T_e^{-3/2}$ is suppressed and laser energy deposition is decoupled from the optimal emission zone.
Conversely, at shorter wavelengths, the higher critical density places the absorption region in a denser plasma, requiring higher laser intensities to heat the larger electron population to the optimal temperature.
Second, the optical depth $\tau_\nu$ must be maintained near unity for efficient extraction of EUV light from inside of the laser-produced tin plasma.
At longer wavelengths, the lower density requires a larger scale length to build up sufficient optical depth.
Since the optimal electron temperature for efficient EUV emission is constrained to 30--40~eV,\cite{Nishihara2008} the plasma sound speed $c_s \propto \sqrt{T_e}$ is effectively fixed.
Therefore, longer pulse durations and larger target radii are required.
Conversely, at shorter wavelengths, laser absorption occurs at higher plasma densities, and the optical depth must be suppressed to near unity.
Shorter pulse durations limit the growth of the plasma scale length, while smaller target radii promote rapid rarefaction through spherical expansion, both acting to prevent excessive optical depth and the associated self-absorption losses.

Figure~\ref{fig:profile} shows the spatial profiles of (a,d,g) the plasma parameters and (b,e,h) the emission and absorption parameters, and (c,f,i) time-integrated and peak normalized emission spectra from 6 to 20~nm at the time of the laser pulse peak for 2.0 $\mu$m, 5.5 $\mu$m, and 10.5 $\mu$m along the CE locus in Fig.~\ref{fig:ce_map}.

Here we focus on the 5.5~$\mu$m case shown in panels (d)--(f), for which the wavelength is 5.5~$\mu$m.
The electron temperature stays at 30--40~eV and the average charge state at 10--13 over several hundred micrometers, which are the optimum temperature and ionization state for emitting in-band EUV radiation.\cite{Sasaki2007,Nishihara2008}
As the guide line indicates, both the laser absorption and the EUV emissivity peak at 19.2~$\mu$m.
At this position, the electron density is $n_\mathrm{e}$ = $1.9 \times 10^{19}$~$\mathrm{cm^{-3}}$, slightly below the critical density $n_{c} = 3.7 \times 10^{19}$~$\mathrm{cm^{-3}}$ at this wavelength, and the optical depth is close to unity.
The spatial profiles of the laser absorption and EUV emissivity decay exponentially with similar scales, as evidenced by their parallel slopes in Fig.~\ref{fig:profile}(e).
This spatial overlap of the laser absorption and EUV in-band emissivity regions within the plasma of unity optical depth is the ideal condition for producing high EUV-CE.

Plasma profiles at other wavelengths along the locus similarly satisfy near-optimal conditions, as shown in Fig.~\ref{fig:profile}(a)--(c) and (g)--(i), consistent with the relatively flat CE profile in Fig.~\ref{fig:ce_map}(b).
Notably, the electron density profiles reflect the wavelength-dependent plasma structure discussed above: at 2~$\mu$m, a steep gradient over a short scale length results from rapid rarefaction of a dense, hot plasma, whereas at 10.5~$\mu$m, a gradual profile extending over a long scale length reflects slow, quasi-one-dimensional expansion sustained by the long pulse and large target.
In both cases the optical depth remains near unity at the position of peak EUV emissivity.
As a consequence, sharp EUV spectra with minimal self-absorption and small out-of-band emission are obtained at all wavelengths as shown in Fig.~\ref{fig:profile}(c), (f), (i), consistent with the relatively flat CE profile in Fig.~\ref{fig:ce_map}(b).
These results suggest that, regardless of the driver wavelength, favorable plasma conditions for efficient EUV emission can be achieved by appropriately optimizing the laser pulse parameters and initial target conditions.
The validity of the one-dimensional spherical approximation is supported by the electron density scale length at $\langle r \rangle_{\mathrm{em}}$ being sufficiently small compared to the initial target size, ensuring that transverse expansion remains negligible relative to the radial direction.
More than 95\% of the simulations yield a ratio below unity, with a median of 0.03 in this parameter survey.

Figure~\ref{fig:ce_map}(h) shows the optimum super-Gaussian order before the pulse peak.
In general, a flat-top temporal pulse shape is preferable for uniform plasma heating.\cite{Mostafa2023}
However, the $2$nd-order super-Gaussian is optimal across most of the wavelength range; the $10$th-order is favored only around 2.5--3.5~$\mu$m, with little CE difference between the two orders at these wavelengths.
This may reflect the absence of a pre-formed plasma in the initial target condition: the slowly rising 2nd-order pulse front allows a plasma density scale length in the under-dense region to develop before the main heating phase, playing a role analogous to target pre-shaping.
This result supports the importance of target shaping, while also suggesting that pulse shape control alone can provide the necessary density scale length to optimize EUV emission.

\begin{figure}
\centering
    \includegraphics[width=0.45\textwidth]{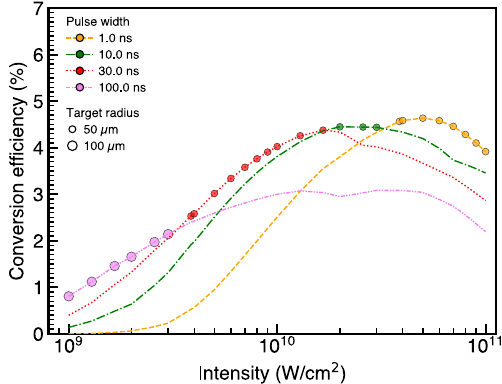}
\caption{Laser intensity dependence of the CE for a 2~$\mu$m drive laser.
The envelope of the maximum CE for each pulse width is presented.
Marker color and line style represent the pulse width: 1.0~ns (dashed, orange), 10.0~ns (dash-dotted, green), 30.0~ns (dotted, red), and 100.0~ns (dash-dot-dotted, pink).
Marker size corresponds to the target radius.}
\label{fig:2um}
\end{figure}

The parameter survey identified the optimum conditions for a 2~$\mu$m driver as an intensity of $5.0 \times 10^{10}$~W/cm$^{2}$, a pulse width of 1~ns, and a target radius of 50~$\mu$m, yielding a maximum CE of 4.64\%.
However, the operation parameter range depends strongly on the type of 2~$\mu$m laser, namely solid-state rod, thin-disk, or fiber architectures.
Therefore, identifying practically accessible operating conditions for each laser architecture is essential for establishing an engineering roadmap toward mid-infrared-driven EUV light sources beyond the optimum parameters alone.

Figure~\ref{fig:2um} shows the envelopes of CE at each pulse width as a function of laser intensity for a 2~$\mu$m driver.
The envelopes reveal that the pulse width yielding the highest CE at an intensity decreases from 100~ns at low intensities to 1~ns at the highest intensities.
In the intensity range of $1.3 \times 10^{10}$--$5 \times 10^{10}$~W/cm$^{2}$, the CE varies only from 4.26\% (pulse width of 30~ns) to 4.64\% (pulse width of 1~ns), indicating that competitive performance is maintained over a broad intensity and pulse width range.
Within this intensity window, the electron temperature and average charge state are maintained near their optimums as shown in Fig.~\ref{fig:ce_map} (c) and (d).

For pulse widths of 1--30~ns, the optical depth stays in the optimum range ($\tau \gtrsim 0.5$) across the intensity window, although the mechanism differs at each end.
At the lower intensity end, longer pulses sustain efficient plasma heating, whereas at the higher end, shorter pulses limit scale-length growth and thus prevent excessive self-absorption.
In contrast, the 100~ns pulse produces an excessively long plasma scale length, which drives the optical depth below 0.3 and prevents the CE from reaching competitive values at higher intensities.
This operating window is advantageous for practical implementation with different 2 $\mu$m laser architectures.
The CE values obtained in the present survey are in reasonable agreement with the experimental results of Mostafa \textit{et al.},\cite{Mostafa2023} in which CE values of approximately 4.4 -- 4.6\% were obtained at an intensity of $7 \times 10^{10}$~W/cm$^{2}$ for 11 and 23~ns flat-top pulses with a 2~$\mu$m driver.

\begin{figure}
\centering
    \includegraphics[width=0.45\textwidth]{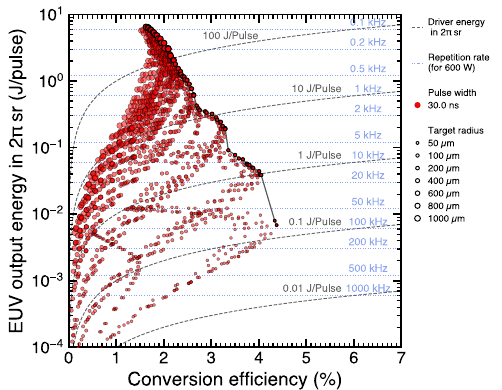}
\caption{Data mapped in the CE--EUV output energy space over 2$\pi$ sr for a pulse width of 30.0 ns and laser wavelength of 2~$\mu$m. The Pareto front is highlighted and connected by a line. Dashed lines indicate the corresponding input energy. Dotted lines represent the repetition rate required to achieve an EUV power of 600 W.}
\label{fig:pareto}
\end{figure}

Not only the CE, but also the absolute EUV output energy should be considered for the lithography application.
Figure~\ref{fig:pareto} maps the simulation results in the CE--EUV output energy space for a pulse width of 30~ns and a laser wavelength of 2~$\mu$m.
To achieve an in-band EUV power of 600~W, comparable to that of current commercial sources,\cite{Zhu2025} possible operating points include, a pulse laser energy of 0.13~J at CE~$= 4.5\%$ with a repetition rate of 100~kHz, or 1~J at CE~$= 4.0\%$ with 15~kHz, with target radii of 50 and 100~$\mu$m, respectively.
This demonstrates that multiple operating points in the CE--EUV output energy space can satisfy the power requirement, offering flexibility in the design of a 2~$\mu$m-driven EUV light source.

In conclusion, a large-scale parameter grid search of more than 140,000 combinations was performed using the one-dimensional radiation-hydrodynamics code \textsc{STAR-1D} to optimize tin-plasma EUV light sources driven by mid-infrared lasers.
The CE map shows a maximum of 5.63\% at 5.5~$\mu$m, with the optimum-intensity locus following $I_{\mathrm{L}}^{\mathrm{opt}} \propto \lambda^{-1.36}$, consistent with previously established scalings.\cite{Nishihara2008,Hemminga2023,Dong2025}
The systematic shift of optimum conditions from long to short wavelengths---longer pulses and larger targets at long wavelengths, shorter pulses and smaller targets at short wavelengths---is well explained by the requirement to maintain the EUV optical depth near unity through plasma scale-length control.
For a 2~$\mu$m driver, the results agree well with experimental data,\cite{Mostafa2023}.
A broad, flat CE profile is obtained over $1 \times 10^{10}$--$5 \times 10^{10}$~W/cm$^{2}$, and the CE--EUV output energy map further identifies multiple operating points capable of achieving 600~W source power, offering practical guidance for 2~$\mu$m-driven EUV source development.

\begin{acknowledgments}
The authors thank the technical support staff for their assistance with computer simulations.
This work was supported by the Japan Science and Technology Agency (JST) K Program Grant Number JPMJKP24M1; the Joint Usage/Research Center Program of the Institute of Laser Engineering (ILE) at The University of Osaka; the ``Power Laser DX Platform'' as shared research equipment under the Ministry of Education, Culture, Sports, Science and Technology (MEXT) Project for Promoting Public Utilization of Advanced Research Infrastructure (Program for Advanced Research Equipment Platforms, Grant No.~JPMXS0450300024); and the Japan Society for the Promotion of Science (JSPS) Core-to-Core Program (Grant No.~JPJSCCA20230003).

\end{acknowledgments}

\section*{Author Declarations}
\subsection*{Conflict of Interest}
The authors have no conflicts to disclose.

\subsection*{Author Contributions}
N.T.: conceptualization (equal), data curation (lead), formal analysis (lead), investigation (lead), methodology (lead), software (equal), validation (lead), visualization (lead), writing/original draft preparation (lead), writing/review \& editing (lead),
Y.Y.: conceptualization (equal), data curation (supporting), formal analysis (supporting), investigation (supporting), methodology (supporting), software (equal), validation (supporting), writing/original draft preparation (supporting), writing/review \& editing (supporting),
A.S.: conceptualization (equal), data curation (supporting), software (equal), supervision (equal),  validation (supporting), writing/original draft preparation (supporting), writing/review \& editing (supporting),
K.N.: conceptualization (equal), formal analysis (supporting), supervision (equal), validation (supporting), writing/original draft preparation (supporting), writing/review \& editing (supporting),
A.S.: software (equal),  supervision (equal), validation (supporting), writing/original draft preparation (supporting), writing/review \& editing (supporting),
T.J.: software (equal),  supervision (equal), validation (supporting), writing/original draft preparation (supporting), writing/review \& editing (supporting),
Y.T.: formal analysis (supporting), validation (supporting),  writing/original draft preparation (supporting), writing/review \& editing (supporting),
K.T.: formal analysis (supporting), validation (supporting), writing/original draft preparation (supporting), writing/review \& editing (supporting),
S.F.: conceptualization (equal), formal analysis (supporting), funding acquisition (equal), project administration (equal), validation (supporting), writing/original draft preparation (supporting), writing/review \& editing (supporting),
M.Y.: conceptualization (equal), funding acquisition (equal), project administration (equal), writing/original draft preparation (supporting), writing/review \& editing (supporting),

\section*{Data Availability Statement}

Complete dataset for Figs.~\ref{fig:ce_map}, \ref{fig:2um}, and \ref{fig:pareto} are available as an interactive tool at \url{https://euvcemap-ileuosaka.streamlit.app/}.
The table data for Fig.~\ref{fig:profile} and part of Fig.~\ref{fig:2um} are openly available in Zenodo at \href{https://doi.org/10.5281/zenodo.20422714}{10.5281/zenodo.20422714}.\cite{TanakaZenodo2026a}
Other data that support the findings of this study are available from the corresponding author upon reasonable request.

\bibliography{APL2026.bib}% Produces the bibliography via BibTeX.

\end{document}